\documentclass[reprint,showkeys,nofootinbib,amsmath,amssymb,aps,prd,]{revtex4-1}
\usepackage[marginal]{footmisc}
\usepackage{graphicx}
\usepackage{dcolumn}
\usepackage{bm}
\usepackage[colorlinks,allcolors=blue,CJKbookmarks=True]{hyperref}
\usepackage{amsmath}
\usepackage{amssymb}
\usepackage{amsfonts}
\usepackage{amssymb}
\usepackage{dcolumn}
\usepackage{bm}
\usepackage{graphicx}
\usepackage{xcolor}
\usepackage{array}
\usepackage{verbatim}
\usepackage{hyperref}
\newcommand{\dd}{\mathrm{d}}

\newcommand{\vk}{\mathbf{k}}
\newcommand{\vp}{\mathbf{p}}

\newcommand{\ba}{\begin{eqnarray}}
	\newcommand{\ea}{\end{eqnarray}}

\begin{document}
	\title{Tensor bispectrum mediated by an excited scalar field during inflation}
	\author{Zhi-Zhang Peng$^{1,2,3}$}
	\email{pengzhizhang@bnu.edu.cn}
	
	\author{Cheng-Jun Fang$^{2,3}$}
	\email{fangchengjun@itp.ac.cn}

	\author{Zong-Kuan Guo$^{2,3,4}$}
	\email{guozk@itp.ac.cn}
	
	\affiliation{$^1$School of Physics and Astronomy, Beijing Normal University, Beijing 100875, People's Republic of China}
	
	\affiliation{$^2$CAS Key Laboratory of Theoretical Physics, Institute of Theoretical Physics, Chinese Academy of Sciences, P.O. Box 2735, Beijing 100190, China}
	
	\affiliation{$^3$School of Physical Sciences, University of Chinese Academy of Sciences, No.19A Yuquan Road, Beijing 100049, China}

	\affiliation{$^4$School of Fundamental Physics and Mathematical Sciences, Hangzhou Institute for Advanced Study, University of Chinese Academy of Sciences, Hangzhou 310024, China}
	
	\begin{abstract}
		We calculate the tensor bispectrum mediated by an excited scalar field during inflation
		and find that the bispectrum peaks in the squeezed configuration,
		which is different from that of gravitational waves induced by enhanced curvature perturbations
		re-entering the horizon in the radiation-dominated era.
		Measuring the bispectrum provides a promising way to distinguish the stochastic gravitational-wave background generated during inflation
		from that generated after inflation.
	\end{abstract}
	\maketitle

	\section{Introduction}
	Recently, several pulsar timing array collaborations,
	including NANOGrav \cite{NANOGrav:2023hde,NANOGrav:2023gor}, PPTA \cite{Zic:2023gta, Reardon:2023gzh},
	EPTA \cite{EPTA:2023fyk,EPTA:2023sfo} and CPTA \cite{Xu:2023wog},
	have individually reported the first compelling evidence for a stochastic gravitational-wave background (SGWB) signal.
	Such a SGWB signal is expected to arise from astrophysical sources or cosmological sources.
	Due to a mild tension between the astrophysical prediction of the spectral shape and the reconstructed one from observed data,
	cosmological sources fit current data better than astrophysical sources.
	If the signal is of cosmological origin, this raises a question:
	was the gravitational wave (GW) signal generated during inflation or after inflation?

	For a Gaussian SGWB, regardless of its astrophysical or cosmological origin,
	the statistical properties of GWs is described only by the power spectrum,
	which depends on the generation mechanism of GWs.
	Distinguishing such a SGWB signal from various sources relies on detailed study of the power spectrum shape.
	Unfortunately, current pulsar timing array observations provide only weak constraints on the shape of the power spectrum,
	so that it is hard to identify the origin of the observed signal~\cite{NANOGrav:2023hvm}.
	Actually, the interactions of quantum fields during inflation result in a large amount of non-Gaussianity of the SGWB.
	In this case, measurements of the power spectrum alone have limited potential in revealing the interactions during inflation.
	Compared to the power spectrum,
	the tensor bispectrum provides richer physical information
	and thus is expected to break the degeneracy of the GW sources.

	In this paper, we calculate for the first time the one-loop tensor bispectrum mediated by an excited scalar field during inflation.
	It is known that the enhancement of the scalar field perturbation during inflation
	generically gives rise to two SGWBs~\cite{Cai:2018tuh,Cai:2019jah,Cai:2019bmk,Zhou:2020kkf,Peng:2021zon,Cai:2021wzd,Inomata:2022ydj}.
	One is sourced by the enhanced scalar field perturbation and stretched to super-horizon scales during inflation.
	The other is produced when curvature perturbations enhanced during inflation re-enter the horizon in the radiation-dominated era.
	The relation for the peak amplitudes and peak frequencies in the spectrum of these two SGWBs
	is discussed in the specific models~\cite{Cai:2019jah,Peng:2021zon} (see \cite{Fumagalli:2021mpc} for general discussion).
	Since the purpose of this work is to calculate the tensor bispectrum in the presence of the enhanced scalar field perturbation,
	we do not specify the mechanism to amplify the scalar field fluctuation
	but consider an exponential amplification of scalar modes during a short period of time.

	Using the in-in formalism~\cite{Maldacena:2002vr,Weinberg:2005vy},
	we compute  
	the contribution of the one-loop diagrams, including the bubble and triangle diagrams, to the tensor bispectrum
	due to the interaction between the scalar field and tensor perturbations.
	Although the one-loop contribution is suppressed by a factor of $H^2/M_{\rm p}^2$
	where $H$ is the Hubble parameter and $M_{\rm p}$ is the reduced Planck mass,
	the large enhancement of the scalar field perturbation can enable the loop contribution to be comparable to the tree-level (one-loop tensor power spectrum from an excited scalar field was first calculated in Refs.~\cite{Ota:2022hvh,Ota:2022xni}, which also pointed out the implied large tensor non-Gaussianity.).
	Therefore, in this paper we focus on the calculation of the one-loop tensor bispectrum.
	Our results show that the tensor bispectrum mediated by an excited scalar field during inflation peaks in a squeezed configuration.
	Actually, the addition of the tree-level contribution to the bispectrum does not change this conclusion~\cite{Maldacena:2002vr,Maldacena:2011nz,Gao:2011vs}.
	Such a shape of the bispectrum is different from
	that of the SGWB induced by enhanced curvature perturbations re-entering the horizon in the radiation-dominated era.
	It is found that the bispectrum of the latter is dominated by the equilateral configuration
	because the source of GWs is composed by gradients of curvature perturbations when re-entering the horizon~\cite{Bartolo:2018evs,Bartolo:2018rku}.
	Measurements of the shape of the tensor bispectrum provide a promising way
	to distinguish the GW signal generated during inflation from that generated after inflation.

	\section{higher order action}
	To expand the action to higher order in tensor perturbations,
	it is convenient to write the metric in the ADM form
	\begin{align}
		ds^2=-N^2d\tau^2+\gamma_{ij}(dx^i+N^id\tau)(dx^j+N^jd\tau),
	\end{align}
	where $N$ and $N^i$ are the lapse function and shift vector, respectively, serving as Lagrange multipliers,
	$\tau$ is the conformal time, and $\gamma_{ij}$ is the spatial components of the metric.
	Following  Maldecena \cite{Maldacena:2002vr},
	we choose the following gauge to fix time and spatial reparametrizations around a spatially-flat Friedmann-Robertson-Walker metric,
	such that $N=a$, $N^i=0$, and
	\ba
	\gamma_{ij} = a^2(\delta_{ij}+h_{ij}+\frac{1}{2}h_{i}^{\;k} h_{kj}+\frac{1}{6}h_{i}^{\;k}h_{k}^{\;l}h_{lj}+\cdots),
	\ea
	where $h_{ij}$ are tensor perturbations which are transverse ($\partial^i h_{ij}=0$) and traceless ($\delta^{ij}h_{ij}=0$).

	We consider the loop contribution to the tensor bispectrum due to a minimally-coupled spectator scalar field $\chi$.
	The scalar field fluctuation is denoted by $\delta \chi$.
	Then the action up to fourth order in $h_{ij}$ and $\delta \chi$ is expanded as
	\begin{align}
		S=&M_{\rm p}^2\int d\tau d^3xa^2 \left(\frac{1}{4}h^{ik}h^{jl}-\frac{1}{8}h^{ij}h^{kl}\right) \partial_k\partial_lh_{ij}   \notag \\
		+&\int d\tau d^3xa^2\left(-\frac{1}{2}h^{ij} + \frac{1}{4}h^{ik} h_{k}{}^{j}\right) \partial_i \delta \chi \partial_j \delta \chi.\label{action2}
	\end{align}
	Then, Legendre transformation gives the interaction Hamiltonian
	\begin{align}
		H_{\rm int}= &M_{\rm p}^2\int d^3x  a^2 \left(-\frac{1}{4}h^{ik}h^{jl}+\frac{1}{8}h^{ij}h^{kl}\right) \partial_k\partial_lh_{ij} \notag \\
		+&\int d^3x  a^2\left(\frac{1}{2}h^{ij} - \frac{1}{4}h^{ik} h_{k}{}^{j}\right) \partial_i \delta \chi \partial_j \delta \chi.
		\label{Ham1}
	\end{align}
	Note that the terms in the first line in Eq.~\eqref{Ham1} represent the classical third-order gravitational Hamitonian,
	which results in the tensor bispectrum dominated by the squeezed configuration~\cite{Maldacena:2002vr,Maldacena:2011nz,Gao:2011vs}.
	The rest in Eq.~\eqref{Ham1} represent the interaction between $\delta \chi$ and $h_{ij}$
	which arise from the kinetic term of the minimally-coupled scalar field in the action.
	We shall focus on such a tensor-scalar interaction and calculate the tensor bispectrum mediated by the scalar field perturbation.

	The Fourier modes of the scalar field perturbation and tensor perturbations are expressed as
	\begin{align}
		\delta  \chi(\tau, \mathbf x)  &= \int \frac{d^3 k}{ (2\pi)^{3}}e^{i\vk\cdot \mathbf x}\delta  \chi_{\vk}(\tau),
		\label{chi:ft}
		\\	
		h_{ij}(\tau, \mathbf x)  &= \int \frac{d^3 k}{ (2\pi)^{3}}e^{i\vk\cdot \mathbf x} \sum_{s=\pm 2}e_{ij}^s(\vk)h^s_{\vk}(\tau),
		\label{h:ft}
	\end{align}
	where  $e_{ij}^{s}(\vk)$ is the polarization tensor with the helicity states $s=\pm 2$,
	satisfying $e_{ij}^{s_1}(\vk)  e^{ij,s_2 *} (\vk) = \delta^{s_1s_2}$ and
	$e_{ij}^{s}(-\vk)=e_{ij}^{s*}(\vk)$.
	The quantized field operators are expanded into the creation and annihilation operators as
	$\delta \chi_{\vk}(\tau)  =u_k(\tau)a_{\vk}+ u^*_k(\tau) a^\dagger_{-\vk}$ and $
	h^{s}_{\vk}(\tau) =v_k(\tau) b^{s}_{\vk}+ v^*_k(\tau) b^{s\dagger}_{-\vk}$.
	The non-vanishing commutation relations for the creation and annihilation operators are given by
	$[a_{\vk_1},a^\dagger_{-\vk_2}] = (2\pi)^3\delta(\vk_1+\vk_2)$ and
	$[b^{s_1}_{\vk_1},b^{s_2 \dagger}_{-\vk_2}] = (2\pi)^3\delta^{s_1 s_2}\delta(\vk_1+\vk_2).$
	The initial conditions for the mode functions, $u_k$ and $v_k$, correspond to the Bunch-Davis vacuum, which are given by
	\begin{align}
		u^{\rm BD}_k(\tau) &= \frac{H}{\sqrt{2k^3}}(1+ik\tau)e^{-ik\tau},\label{defuq}\\
		v^{\rm BD}_k(\tau) &= \frac{2H}{M_{\rm p}\sqrt{2k^3}}(1+ik\tau)e^{-ik\tau}.\label{defvq}
	\end{align}

	\section{Tensor bispectrum}
\begin{figure}[t!]
		\includegraphics[width=0.48\textwidth]{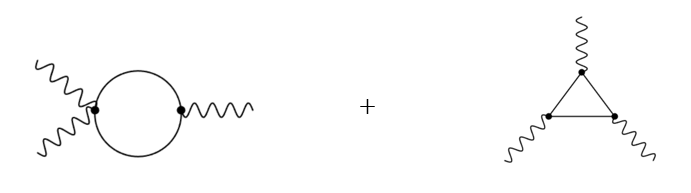}
		\caption{One-loop Feynman diagrams
			including the bubble ({\it left}) and triangle ({\it right}) diagrams.}
		\label{fyn}
	\end{figure}

	Using the in-in formalism~\cite{Maldacena:2002vr,Weinberg:2005vy},
	we compute the one-loop contribution to the tensor bispectrum from the gravitational interaction with the excited scalar field.
	The in-in formalism was firstly proposed by Maldacena to compute the non-Gaussianity of primordial tensor perturbations in the context of single-field slow-roll inflation~\cite{Maldacena:2002vr,Maldacena:2011nz} and was later applied to extra fields \cite{Dimastrogiovanni:2018gkl,Goon:2018fyu}, non-attractor phase for tensor fluctuations \cite{Ozsoy:2019slf},
	massive gravity theory \cite{Fujita:2019tov},
	generalized G-inflation \cite{Gao:2011vs,Gao:2012ib,Choudhury:2012whm,Huang:2013epa},
	$\alpha$-vacuum \cite{Kanno:2022mkx,Gong:2023kpe}, axion-gauge field models \cite{Agrawal:2017awz,Agrawal:2018mrg}, and more generally, effective field theory \cite{Naskar:2018rmu,Naskar:2019shl,Bordin:2020eui,Cabass:2022jda} .
	The equal-time correlators are computed with the in-in formalism via the following formula
	\begin{align}
		\langle \mathcal{O}\rangle  = &\lim_{\tau_0\to -\infty(1-i\epsilon)} \langle 0 | \bar T \exp\left(
		i \int^\tau_{\tau_0^*}d\tau' H_{{\rm int},I}(\tau')
		\right)  \notag\\
		& \times \mathcal{O}_I(\tau) T \exp\left(
		-i \int^\tau_{\tau_0}d\tau'' H_{{\rm int},I}(\tau'')
		\right) |0\rangle,\label{eq45}
	\end{align}
	where the subscript $I$ labels fields in the interaction picture,
	$T$ and $\bar T$ denote the time and anti-time ordering operator, respectively.
	According to the order counting parameter, 
	Eq.~\eqref{eq45} is expanded to
	\ba
	\langle \mathcal{O}\rangle_0&=&\langle 0|\mathcal{O}_{I}(\tau)|0\rangle,	
	\label{O0}		\\
	\langle \mathcal{O}\rangle_1&=&2\,\text{Im} \int^\tau_{\tau_0}d\tau' \langle 0|\mathcal{O}_I(\tau) H_{{\rm int},I}(\tau')|0\rangle,
	\label{O1}		\\
	\langle \mathcal{O}\rangle_{2a}&=&  \int^\tau_{\tau_0^*}d\tau' \int^\tau_{\tau_0}d\tau'' \nonumber \\
	&&\times    \langle 0|  H_{{\rm int},I}(\tau') \mathcal{O}_I(\tau) H_{{\rm int},I}(\tau'')	|0\rangle,
	\label{O2a}      \\
	\langle \mathcal{O}\rangle_{2b}&=& - 2\,\text{Re}   \int^\tau_{\tau_0}d\tau'\int^{\tau'}_{\tau_0}d\tau'' \nonumber \\
	&&\times \langle 0| \mathcal{O}_I(\tau)  H_{{\rm int},I}(\tau') H_{{\rm int},I}(\tau'')|0\rangle,
	\label{O2b}\\
	\langle \mathcal{O}\rangle_{3a}&=&2\, \text{Im} \int^\tau_{\tau_0}d\tau'\int^{\tau}_{\tau_0}d\tau'' \int^{\tau''}_{\tau_0}d\tau''' \nonumber \\
	&&\times \langle 0|  H_{{\rm int},I}(\tau')\mathcal{O}_I(\tau)  H_{{\rm int},I}(\tau'')H_{{\rm int},I}(\tau''')|0\rangle,
	\label{O3a} \\
	\langle \mathcal{O}\rangle_{3b}&=& -2\,\text{Im}  \int^\tau_{\tau_0^*}d\tau' \int^{\tau'}_{\tau_0}d\tau''  \int^{\tau''}_{\tau_0}d\tau''' \nonumber \\
	&&\times    \langle 0| \mathcal{O}_I(\tau) H_{{\rm int},I}(\tau') H_{{\rm int},I}(\tau'')H_{{\rm int},I}(\tau''')	|0\rangle ,
	\label{O3b}
	\ea
	with $\langle \mathcal{O}\rangle_2\equiv \langle \mathcal{O}\rangle_{2a}+\langle \mathcal{O}\rangle_{2b}$ and
	$\langle \mathcal{O}\rangle_3\equiv \langle \mathcal{O}\rangle_{3a}+\langle \mathcal{O}\rangle_{3b}$.
	Since the vacuum expectation values of $\mathcal{O}$ are evaluated at the end of inflation,
	we set $\tau=0$ in the end of calculation.
	For the three-point correlators of GWs, $\langle \mathcal{O}\rangle_{0}$ vanishes.
	Given the interaction Hamiltonian~\eqref{Ham1},
	we can calculate the vacuum expectation value of the three-point correlators $\langle \mathcal{O}\rangle_{1}$,
	$\langle \mathcal{O}\rangle_{2}$ and $\langle \mathcal{O}\rangle_{3}$.
	The first one is the tree-level contribution
	while $\langle \mathcal{O}\rangle_{2}$ and $\langle \mathcal{O}\rangle_{3}$
	are the one-loop contributions,
	which correspond to the bubble (left) and triangle (right) Feynman diagrams in Fig.~\ref{fyn}, respectively.
	Eqs.~\eqref{O2a}-\eqref{O3b} are the basic equations we use in what follows.

	Traditionally, to capture the shape of the tensor bispectrum it is convenient to define the shape function $S^{s_1s_2s_3}(k_1,k_2,k_3)$ by
	\begin{align}
		&\left\langle h^{s_1}_{\mathbf k_1}(\tau) h^{s_2}_{\mathbf k_2}(\tau)h^{s_3}_{\mathbf k_3}(\tau) \right\rangle
		= (2\pi)^3 \delta(\mathbf{k_1+k_2+k_3}) \notag \\
		&\qquad \times \frac{\mathcal{P}_h^2}{k_1^2k_2^2k_3^2}S^{s_1s_2s_3}(k_1,k_2,k_3),
	\end{align}
	where $\mathcal{P}_h$ is the power spectrum of tensor perturbations
	given by $\mathcal{P}_h=\frac{2H^2}{\pi^2M^2_{\rm p}}$ in the slow-roll inflationary model.
	The amplitude of the bispectrum relies on the amplification factor of the scalar field perturbation.
	In this work, we are interested in the shape of the bispectrum rather than its amplitude.
	Since the one-loop contribution is suppressed by a factor of $H^2/M_{\rm p}^2$,
	we rescale the shape function as
	$\left\langle h^{s_1}_{\mathbf k_1} h^{s_2}_{\mathbf k_2} h^{s_3}_{\mathbf k_3} \right\rangle'(H/M_{\rm p})^{-6}k^2_1k^2_2k^2_3$,
	where the prime denotes omitting the delta function for the momentum conservation.

	\section{One-loop contributions from the excited scalar field}
	We consider 
	an exponential growth of the scalar field perturbation on sub-horizon scales,
	which can be achieved through parametric resonance~\cite{Cai:2018tuh,Cai:2019jah,Cai:2019bmk,Zhou:2020kkf,Peng:2021zon,Cai:2021wzd,Inomata:2022ydj}.
	The amplification of $\delta \chi$ from $t_i$ to $t_f$ is given by
	$\delta\chi=e^{\mu H(t_f-t_i)}\delta\chi_{\rm BD}$,
	where $\mu$ is a dimensionless constant.
	In terms of the conformal time, the amplification factor $\mathcal{A} = e^{\mu H(t_f-t_i)}$ can be written as
	$\mathcal{A} =\left(\tau_i/\tau_f\right)^\mu$,
	where $\tau_i$ and $\tau_f$ correspond to $t_i$ and $t_f$, respectively.
	We assume the amplification factor is constant when $\tau > \tau_f$.
	Therefore, 
	the amplification factor as a function with $\tau$ is parameterized by
	\begin{align}
		\mathcal{A}(\tau) =
		\begin{cases}
			1 &\tau<\tau_i,\\
			\left(\frac{\tau_i}{\tau}\right)^\mu &\tau_i\le\tau\le\tau_f, \\
			\left(\frac{\tau_i}{\tau_f}\right)^\mu &\tau > \tau_f.	
		\end{cases}
		\label{defxi}
	\end{align}
	In practise, we set $\mathcal{A}=0$ when $\tau < \tau_i$ so that the vacuum contribution is subtracted.
	For simplicity, we consider that the scalar field perturbation is amplified only at $k=k_*$
	($k_*$ is the pivot scale)
	and tensor perturbations remain unchanged.
	Thus the mode functions are given by
	\begin{align}\label{evoldel}
		u_k(\tau)=&\mathcal{A}(\tau)u^{\rm BD}_k(\tau)\delta(\ln k/k_*),\\
		v_k(\tau)=&v^{\rm BD}_k(\tau).\label{hevoldel}
	\end{align}
	In principle it is straightforward to generalize the delta function to a realistic momentum distribution.


	Now let us consider the one-loop contribution to the tensor bispectrum from the bubble diagram in Fig.~\ref{fyn}.
	For the polarization configuration $++$$+$, the results are
	\ba\label{h2a}
	&& \left\langle h^{+}_{\mathbf k_1}(\tau) h^{+}_{\mathbf k_2}(\tau)h^{+}_{\mathbf k_3}(\tau) \right\rangle_{2a}'
	\notag
	=
	-\frac{1}{2}\left(\frac{H}{M_{\rm p}}\right)^6\Theta_{2-\tilde{k}_1}	\bar{w}^{+++}_{k_1} \notag\\
	&& \quad \times \;  \frac{1}{k_1^3k_2^3k_3^3 k_*^4} \int^0_{x_0}dx'  \int^0_{x_0}dx''
	\mathcal{F}(\tilde{k}_1,x')\mathcal{G}(\tilde{k}_2,\tilde{k}_3,x'') \nonumber \\
	&& \quad  +\;2\,\text{perms}, \\
	&& \left\langle h^{+}_{\mathbf k_1}(\tau) h^{+}_{\mathbf k_2}(\tau)h^{+}_{\mathbf k_3}(\tau) \right\rangle_{2b}'
	=\frac{1}{2}{\rm Re}\left(\frac{H}{M_{\rm p}}\right)^6\Theta_{2-\tilde{k}_1}	\bar{w}^{+++}_{k_1} \notag \\
	&& \quad \times \; \frac{1}{k_1^3k_2^3k_3^3 k_*^4} \int^0_{x_0}dx'
	\int^{\tau'}_{x_0}dx''\Big[\mathcal{F}(-\tilde{k}_1,x')\mathcal{G}(\tilde{k}_2,\tilde{k}_3,x'') \nonumber \\
	&& \quad +\;\mathcal{F}^*(\tilde{k}_1,x'')\mathcal{G}^*(-\tilde{k}_2,-\tilde{k}_3,x')\Big]+2\,\text{perms},
	\ea
	where $x=k_* \tau$, $\tilde{k}_i=k_i/k_*$,
	and $\Theta_{2-\tilde{k}_1}$ is the Heaviside step function with the argument $ 2-\tilde{k}_1$ that implies the momentum conservation.
	Here we have introduced
	\ba
	\mathcal{F}(\tilde{k}_1,x) &=& \frac{1}{x^2}  (1+i\tilde{k}_1x)(1+ix)^2 e^{-i(\tilde{k}_1+2)x}\mathcal{A}(x/k_*)^2, \nonumber \\
	\mathcal{G}(\tilde{k}_2,\tilde{k}_3,x) &=& \frac{1}{x^2} (1-i\tilde{k}_2x)(1-i\tilde{k}_3x)(1-ix)^2 \nonumber \\
	&& \times \;e^{i(\tilde{k}_2+\tilde{k}_3+2)x}	\mathcal{A}(x/k_*)^2, \nonumber \\
	\bar{w}^{+++}_{k_1}&=&\frac{- A_{k_1k_2k_3}^2k_T^2k_*^4(k_1^2-4k_*^2)^2}{512\pi^2k_1^3k_2^2k_3^2},
	\ea
	where  $A_{k_1k_2k_3}\equiv1/4\sqrt{k_T(k_T-2k_3)(k_T-2k_2)(k_T-2k_1)}$ is the area of the triangle of sides $k_i$ and $k_T\equiv k_1+k_2+k_3$. The detailed derivation of the one-loop contribution from the bubble diagram  is given in Appendix \ref{A}.

\begin{figure}[t!]
		\includegraphics[width=0.35\textwidth]{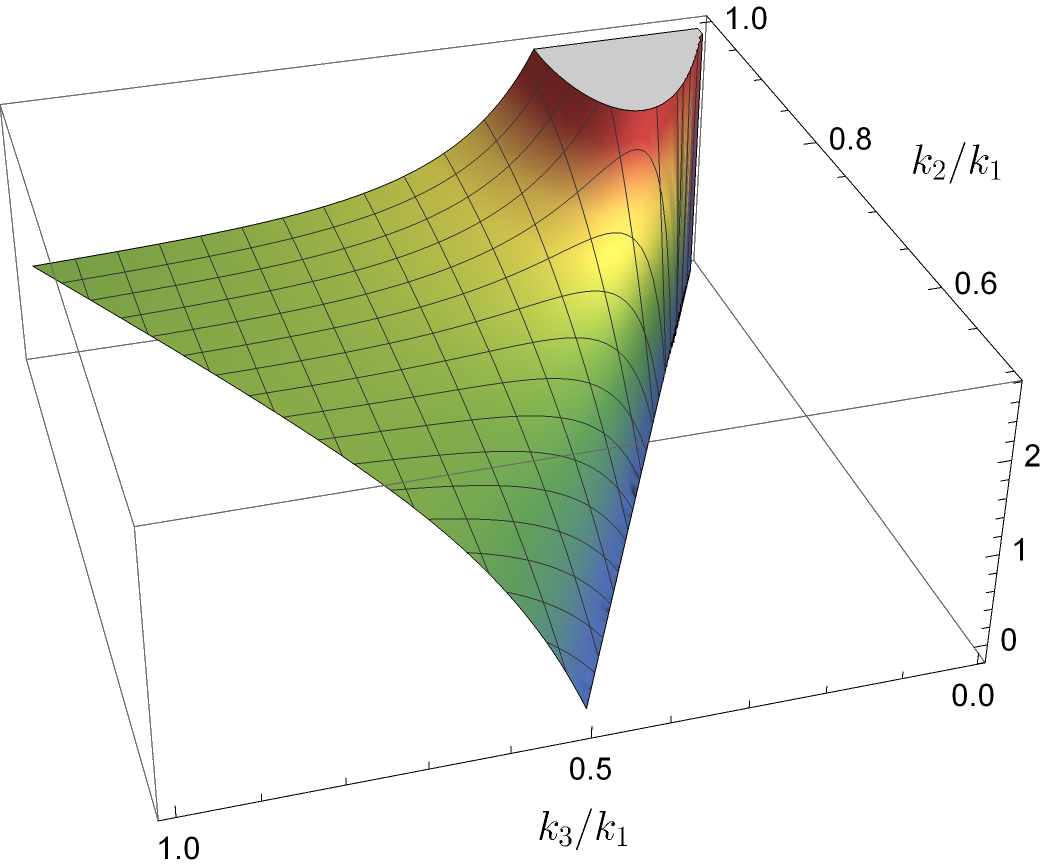}
		\caption{Rescaled shape function for the $+$+$+$ polarization from the bubble diagram, where we choose $k_1/k_*=0.1$.
			The plot is normalized to unity for equilateral configurations $k_2/k_1 = k_3/k_1 = 1$.}
		\label{bubble}
	\end{figure}

	The scalar field perturbation is enhanced on sub-horizon scales due to parametric resonance.
	Without loss of generality,
	we set the initial time and final time of amplification as $x_i=-100$ and $x_f=-10$,
	and set the index as $\mu=2$.
	This implies that $\delta \chi$ is amplified by a factor of $(x_i/x_f)^\mu=10^2$ within the horizon.
	The rescaled shape function $\langle h^{+}_{\mathbf k_1} h^{+}_{\mathbf k_2} h^{+}_{\mathbf k_3} \rangle'(H/M_{\rm p})^{-6}k^2_1k^2_2k^2_3$
	is shown in Fig.~\ref{bubble}.
	We see that the tensor bispectrum peaks in the squeezed configuration.
	As discussed in Refs.~\cite{Ota:2022hvh,Ota:2022xni},
	for the power spectrum of tensor perturbations
	the delta function in~\eqref{evoldel} violates the causality resulting in the infrared divergence.
	To overcome such an issue, a log-normal distribution with a finite width $\Delta$ is introduced to replace the delta function,
	\begin{align}\label{log-normal}
		\delta(\ln k/k_*)\to\frac{1}{\sqrt{2\pi}\Delta}e^{-\frac{\ln(k/k_*)^2}{2\Delta^2}}.
	\end{align}
	Although our work is free from this issue,
	we also consider the tensor bispectrum in the log-normal case.
	We find that the log-normal function gives the same shape of the bispectrum as the delta function.
	Compared to the delta function case,
	the value of the shape function in the squeezed limit is smaller in the log-normal case.
	The reason is  as follows.
	For the log-normal distribution, as $k \to 0$, the step function behaves as a linear function of $k$,
	resulting in the peak value 
	smaller than one in the delta function case. The full comparison between delta case and log-normal case are shown in Appendix \ref{B}.
	Since this work focuses on the shape of the tensor bispectrum,
	in what follows we shall consider only the delta function. 

	Similar to the bubble diagram, it is straightforward to calculate the one-loop contribution from the triangle diagram.	
	From Eqs.~\eqref{O3a} and \eqref{O3b} we have	
	\ba
	&& \left\langle h^{s_1}_{\mathbf k_1}(\tau) h^{s_2}_{\mathbf k_2}(\tau)h^{s_3}_{\mathbf k_3}(\tau) \right\rangle'_{3a} =
	\left(\frac{H}{M_{\rm p}}\right)^6\Theta_{2-\tilde{k}_1}	\frac{4}{k_1^3k_2^3k_3^3 k_*} \nonumber \\
	&& \quad \times \left(16 A_{k_1k_2k_3}^2-\frac{k^2_1 k^2_2 k^2_3}{k^2_*}\right)^{-1/2}\mathcal{D}^{s_1s_2s_3}(k_1,k_2,k_3) \nonumber \\
	&& \quad \times \;\text{Im}\int^0_{x_0^*}dx'  \int^0_{x_0}dx''\int^{x''}_{x_0^*}dx''' 	\mathcal{X}(\tilde{k}_1,x') \nonumber \\
	&& \quad \times \left[\mathcal{Y}(\tilde{k}_2,x'')\mathcal{Z}(\tilde{k}_3,x''')
	+\mathcal{Y}(\tilde{k}_3,x'')\mathcal{Z}(\tilde{k}_2,x''')\right] \nonumber \\
	&& \quad +\;2\,\text{perms}\,,
	\\
	&& \left\langle h^{s_1}_{\mathbf k_1}(\tau) h^{s_2}_{\mathbf k_2}(\tau)h^{s_3}_{\mathbf k_3}(\tau) \right\rangle'_{3b} =
	\left(\frac{H}{M_{\rm p}}\right)^6\Theta_{2-\tilde{k}_1}	\frac{4}{k_1^3k_2^3k_3^3 k_*} \notag \\
	&& \quad \times \left(16 A_{k_1k_2k_3}^2-\frac{k^2_1 k^2_2 k^2_3}{k^2_*}\right)^{-1/2}\mathcal{D}^{s_1s_2s_3}(k_1,k_2,k_3) \nonumber \\
	&& \quad \times\;\text{Im}  \int^0_{x_0^*}dx'  \int^{x'}_{x_0}dx''\int^{x''}_{x_0^*}dx''' \mathcal{X}(-\tilde{k}_1,x') \nonumber \\
	&& \quad \times \left[\mathcal{Y}(\tilde{k}_2,x'')\mathcal{Z}(\tilde{k}_3,x''')
	+\mathcal{Y}(\tilde{k}_3,x'')\mathcal{Z}(\tilde{k}_2,x''')\right] \nonumber \\
	&& \quad +\;2\,\text{perms}\,,
	\ea
	where
	\ba
	\mathcal{X}(k,x)&=&\frac{1}{x^2}(1+ikx)(1+ix)^2e^{-i(k+2)x}\mathcal{A}(x/k_*)^2,\nonumber\\
	\mathcal{Y}(k,x)&=&\frac{1}{x^2}(1-ikx)(1+x^2)e^{ikx}\mathcal{A}(x/k_*)^2,\nonumber\\
	\mathcal{Z}(k,x)&=&\frac{1}{x^2}(1-ikx)(1-ix)^2e^{ikx}\mathcal{A}(x/k_*)^2.
	\ea
	Here $\mathcal{D}^{s_1s_2s_3}(k_1,k_2,k_3)$ is the product of projection tensors,
	which can be written as a compact form for the $+$+$+$ polarization in the equilateral situation (i.e., $k_1=k_2=k_3$).
	\ba
	\mathcal{D}^{+++}(k_1,k_1,k_1)=\frac{365k_1^6}{6912}-\frac{61k_1^4k_*^2}{192}+\frac{9k_1^2k_*^4}{16}-\frac{k_*^6}{4}. \nonumber
	\ea

	\begin{figure}[t!]
		\includegraphics[width=0.4\textwidth]{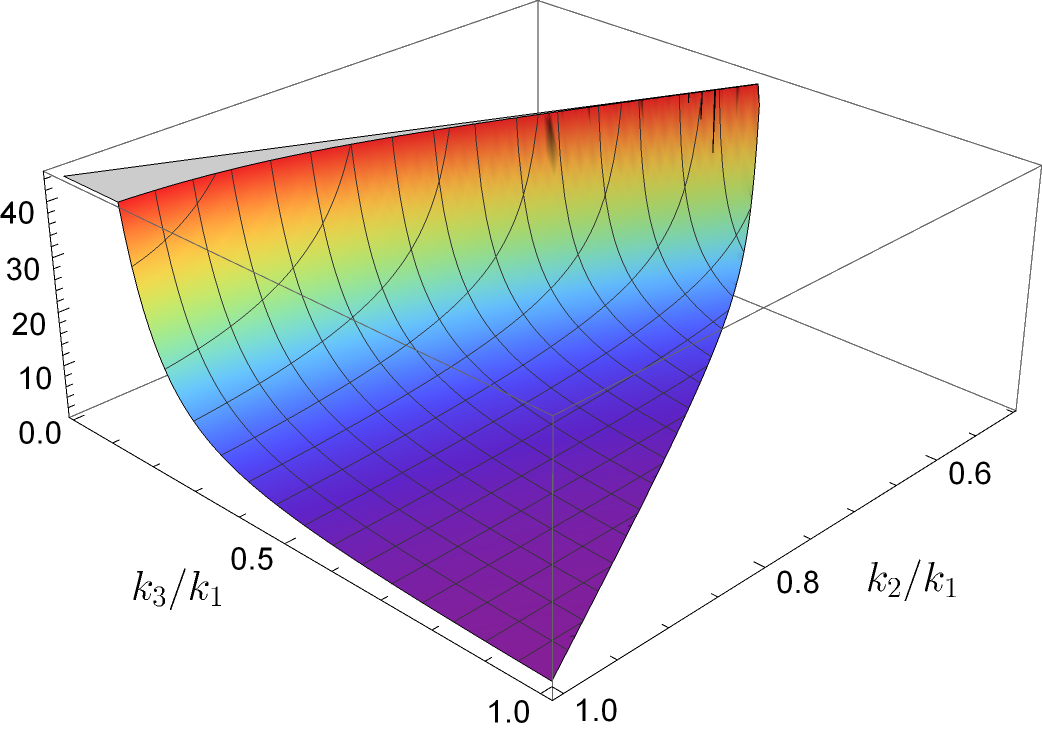}
		\caption{Rescaled shape function for the $+$+$+$ polarization from the triangle diagram, where we choose $k_1/k_*=0.1$.
			The plot is normalized to unity for equilateral configurations $k_2/k_1 = k_3/k_1 = 1$.}
		\label{triangle}
	\end{figure}

	In Fig.~\ref{triangle}, we display the rescaled shape function for the $+$+$+$ polarization in the case of the triangle diagram.
	We can see that the tensor bispectrum peaks in the squeezed configuration.
	Such a result is the same as that obtained in the case of the bubble diagram. 
	The emergence of the squeezed shape appears to resemble the scenario in quasi-single field inflation, where the modes for light isocurvaton survive for a long time on super-horizon scales, thereby leading to a quasi-local shape~\cite{Chen:2009zp,Chen:2010xka}. However, the scalar field perturbation is enhanced on sub-horizon scales in our model. This anomaly could potentially be attributed to the complex interplay between tensor perturbations  and scalar field perturbations during inflation. Considering a specific inflationary model would aid in confirming this result.
	Moreover, we find the peak value of the bispectrum is significantly larger than that from the bubble diagram.
	The reason is the triangle diagram involves more scalar propagators.
	Therefore, in this case the contribution from the triangle diagram dominates the one-loop contributions to the tensor bispectrum.
	
	\begin{figure}[t!]
		\includegraphics[width=0.55\textwidth]{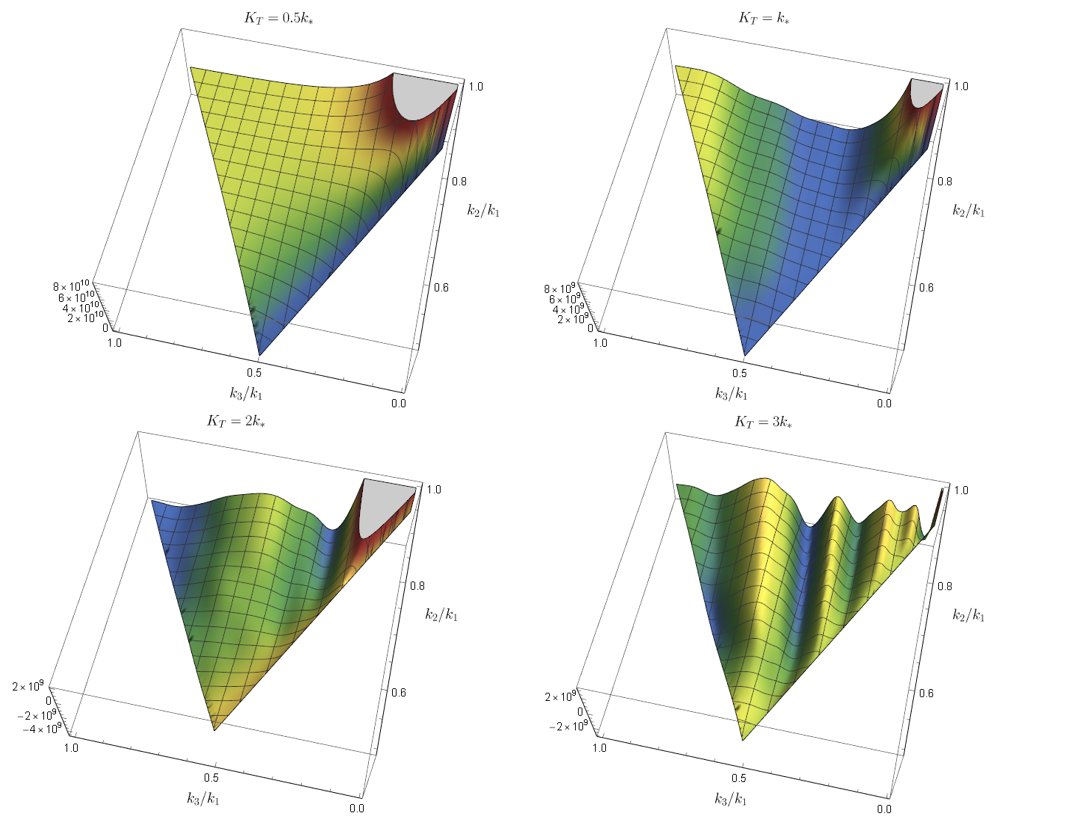}
		\caption{Rescaled shape function for the $+$+$+$ polarization from the bubble diagram, where we set $k_T=0.5k_*$, $k_*$, $2k_*$ and $3k_*$, respectively.}
		\label{running}
	\end{figure}
	We note that the tensor bispectrum is no longer scale-invariant, due to the presence of a pivot scale $k_*$. In this context, an additional critical characteristic, besides its shape, is the running of the bispectrum. We take bubble diagram for example. Following the tradition in \cite{Chen:2010xka}, we show the dependence of the rescaled shape function on the momenta ratio $k_2/k_1$ and $k_3/k_1$, while fixing the perimeter of the momentum triangle $k_T$ in Fig.~\ref{running}. A outstanding feature is that the tensor bispectrum oscillates in k-space. The underlying physics can be readily understood in terms of generation mechanism. The scalar field perturbation experiences a resonant amplification on sub-horizon scales from $\tau_i$ to $\tau_f$, which is similar to resonant non-Gaussianity shown in \cite{Chen:2008wn,Flauger:2010ja}. Moreover, since we have chosen a specific starting point for resonant amplification, the oscillatory period in k-space is approximately equal to $2\pi/x_i$ which is a constant. Therefore, as $k_T$ increases covering multiple periods, the oscillations become increasingly evident.


	\section{Conclusions and discussions}
	We have calculated the tenor bispectrum mediated by an excited scalar field during inflation.
	We consider the one-loop contributions to the tenor bispectrum from the bubble and triangle Feynman diagrams.
	In both cases, the bispectrum peaks in the squeezed configuration.
	After inflation, enhanced scalar perturbations induce another SGWB when re-entering the horizon.
	The tensor bispectrum of the SGWB is dominated by the equilateral configuration.
	Hence measurements of the bispectum provide a potential way to distinguish the SGWBs.

	We consider the interaction between tensor perturbations and the spectator scalar field perturbation
	due to the canonical kinetic term, which is independent of the potential of the scalar field.
	Moreover, our calculation is based on the exponential growth of the scalar field perturbation only for $k=k_*$ mode.
	Such an amplification factor can be realized through parametric resonance.
	For a log-normal momentum distribution around $k_*$,
	our conclusions are unchanged.

	As an illustration, we show the shape function of the bispectrum only for the $+$$+$$+$ polarization in Fig.~\ref{bubble} and Fig.~\ref{triangle}.
	Due to parity symmetry, the bispectrum for the $-$$-$$-$ polarization is the same as that for the $+$$+$$+$ polarization.
	For the $+$$+$$-$ and $-$$-$$+$ polarizations we have checked that the bispectrum takes its maximal value in the squeezed configuration. 
	
	It is pointed out that a large amplification of the scalar field perturbation during inflation
	enables the loop power spectrum to dominate over the tree-level power spectrum in the in-in formalism,
	indicating the breakdown of the perturbation theory~\cite{Inomata:2022yte,Ota:2022hvh,Ota:2022xni}.
	The necessary conditions for the subdominant loop power spectrum are discussed in Ref.~\cite{Inomata:2022yte}.
	In our model, the conditions are related to the final value of the amplification factor and thus are assumed to be satisfied.
	
	It is widely held that tensor bispectrum is strongly suppressed at interferometers scales owing to Shapiro time-delay effects associated with the propagation of GWs~ \cite{Adshead:2009bz,Bartolo:2018evs,Bartolo:2018rku,Kehagias:2024plp}. A typical counterexample is the flattened configuration of the tensor bispectrum, where phase differences from source to detection are eliminated~\cite{Powell:2019kid}. Despite the difficulties in measurement, we argue that we theoretically provide a perspective to distinguish the SGWBs generated during inflation from that generated after inflation. Besides, the quadrupolar anisotropy of SGWBs serves as an indirect probe for squeezed tensor non-Gaussianity, which can escape the suppression by propagation effects~\cite{Dimastrogiovanni:2019bfl}. Examining the quadrupolar anisotropy in our model may be interesting and we leave it for future work.

	Finally, we mention that the SGWB can be sourced by the production of the gauge quanta during inflation
	with the coupling $\phi F\tilde{F}$ of the pesudoscalar field to the gauge field.
	The shape of the tensor bispectrum for the $+$$+$$+$ polarization is very close to equilateral~\cite{Cook:2013xea}.
	Although this shape is the same as that of the SGWB induced by enhanced curvature perturbations in the radiation-dominated era,
	with the help of parity symmetry, we can distinguish these two SGWBs.

	\section{Acknowledgements}
	We thank Xingang Chen, Antonio Riotto, Misao Sasaki and Atsuhisa Ota for comments on the manuscript, and Chao Chen, Zhong-Zhi Xianyu, Yi Wang and Yuhang Zhu for useful discussions. This work is supported
	in part by the National Key Research and Development Program of China Grant No. 2020YFC2201501,
	in part by the National Natural Science Foundation of China under Grant No. 12075297 and No. 12235019.

	\appendix
	
	\section{\label{A}A detailed derivation of the one-loop contribution}
	In this appendix, I provide a detailed derivation of the one-loop contribution from the bubble and triangle diagrams.
The Hamiltonian has both third-order and fourth-order contributions, which results in containing two parts of contributions $\langle \mathcal{O}\rangle_{2a,1}$ and $\langle \mathcal{O}\rangle_{2a,2}$
\begin{widetext}
\begin{align}
	\left\langle h^{s_1}_{\mathbf k_1}(\tau) h^{s_2}_{\mathbf k_2}(\tau)h^{s_3}_{\mathbf k_3}(\tau) \right\rangle_{2a,1}=\int^\tau_{\tau_0^*}d\tau' \int^\tau_{\tau_0}d\tau''\langle 0| H_{\rm int}^{(3)}(\tau')h^{s_1}_{\mathbf k_1}(\tau) h^{s_2}_{\mathbf k_2}(\tau)h^{s_3}_{\mathbf k_3}(\tau)H_{\rm int}^{(4)}(\tau'')|0 \rangle\,,\\
	\left\langle h^{s_1}_{\mathbf k_1}(\tau) h^{s_2}_{\mathbf k_2}(\tau)h^{s_3}_{\mathbf k_3}(\tau) \right\rangle_{2a,2}=\int^\tau_{\tau_0^*}d\tau' \int^\tau_{\tau_0}d\tau''\langle 0| H_{\rm int}^{(4)}(\tau')h^{s_1}_{\mathbf k_1}(\tau) h^{s_2}_{\mathbf k_2}(\tau)h^{s_3}_{\mathbf k_3}(\tau)H_{\rm int}^{(3)}(\tau'')|0 \rangle\,.
\end{align}
\end{widetext}
Based on Eq.~\eqref{chi:ft} and Eq.~\eqref{h:ft}, we can expand the above two equations separately as
\begin{align}\label{2a1}
	&\left\langle h^{s_1}_{\mathbf k_1}(\tau) h^{s_2}_{\mathbf k_2}(\tau)h^{s_3}_{\mathbf k_3}(\tau) \right\rangle_{2a,1}
	\notag \\
	=&
	-\frac{1}{8} \int^\tau_{\tau_0^*}d\tau' a(\tau')^2 \int^\tau_{\tau_0}d\tau''   a(\tau'')^2   \prod_{A=1}^7\left(\int \frac{d^3p_A}{(2\pi)^{3}}\right)
	\notag \\
	&\times(2\pi)^3\delta\left(\sum_{A=1}^3 \mathbf p_A\right)(2\pi)^3\delta\left(\sum_{A=4}^7 \mathbf p_A\right)   
	\notag \\
	&	\times \sum_{s,s_4,s_5} e_{ij}^{s}(\hat p_1) p_{2i}p_{3j}    e_{km}^{s_4}(\hat p_4) e_{ml}^{s_5}(\hat p_5) p_{6k}p_{7l} 
	\notag \\
	&	\times  \langle 0| h^{s}_{\mathbf p_1}(\tau')h^{s_1}_{\mathbf k_1}(\tau)h^{s_2}_{{\mathbf k_2}}(\tau)h^{s_3}_{{\mathbf k_3}}(\tau) h^{s_4}_{\mathbf p_4}(\tau'') h^{s_5}_{\mathbf p_5}(\tau'')|0\rangle
	\notag \\
	&\times \langle 0| \delta \chi_{\mathbf p_2}(\tau')  \delta \chi_{\mathbf p_3} (\tau')  \delta \chi_{\mathbf p_6}(\tau'')  \delta \chi_{\mathbf p_7}(\tau'')	|0\rangle.
\end{align}
and
\begin{align}\label{2a2}
	&\left\langle h^{s_1}_{\mathbf k_1}(\tau) h^{s_2}_{\mathbf k_2}(\tau)h^{s_3}_{\mathbf k_3}(\tau) \right\rangle_{2a,2}
	\notag \\
	=&
	-\frac{1}{8} \int^\tau_{\tau_0^*}d\tau' a(\tau')^2 \int^\tau_{\tau_0}d\tau''   a(\tau'')^2   \prod_{A=1}^7\left(\int \frac{d^3p_A}{(2\pi)^{3}}\right)
	\notag \\
	&\times(2\pi)^3\delta\left(\sum_{A=1}^3 \mathbf p_A\right)(2\pi)^3\delta\left(\sum_{A=4}^7 \mathbf p_A\right)   
	\notag \\
	&	\times \sum_{s,s_4,s_5} e_{ij}^{s_1}(\hat p_1) p_{2i}p_{3j}    e_{km}^{s_4}(\hat p_4) e_{ml}^{s_5}(\hat p_5) p_{6k}p_{7l} 
	\notag \\
	&	\times  \langle 0|  h^{s_4}_{\mathbf p_4}(\tau') h^{s_5}_{\mathbf p_5}(\tau')h^{s_1}_{\mathbf k_1}(\tau)h^{s_2}_{{\mathbf k_2}}(\tau)h^{s_3}_{{\mathbf k_3}}(\tau)  h^{s}_{\mathbf p_1}(\tau'')|0\rangle
	\notag \\
	&\times \langle 0|  \delta \chi_{\mathbf p_6}(\tau')  \delta \chi_{\mathbf p_7}(\tau') \delta \chi_{\mathbf p_2}(\tau'')  \delta \chi_{\mathbf p_3} (\tau'')	|0\rangle	.
\end{align}
We only consider connected graphs, and Eq.~\eqref {2a1} can be simplified as
\begin{align}\label{2a,1s}
	&\left\langle h^{s_1}_{\mathbf k_1}(\tau) h^{s_2}_{\mathbf k_2}(\tau)h^{s_3}_{\mathbf k_3}(\tau) \right\rangle_{2a,1}\notag\\ =&-\frac{1}{2}(2\pi)^3\delta\left(\mathbf k_1+ \mathbf k_2+\mathbf k_3\right) \int^\tau_{\tau_0^*}d\tau' a(\tau')^2  \int^\tau_{\tau_0}d\tau''    a(\tau'')^2 \notag\\
	&\times\prod_{A=2}^3\left(\int \frac{d^3p_A}{(2\pi)^{3}}\right) (2\pi)^3\delta\left(\sum_{A=2}^3 \mathbf p_A-\mathbf k_1\right)
	\notag\\
	 	&\times e_{ij}^{s_1*}(\hat k_1) p_{2i}p_{3j}  
  e_{km}^{s_2*}(\hat k_2) e_{ml}^{s_3*}(\hat k_3) p_{2k}p_{3l}  
	 v^*_{k_1}(\tau) v_{k_2}(\tau) v_{k_3}(\tau) 
	 \notag\\
	 	&\times v_{k_1}(\tau') u_{p_2}(\tau') u_{p_3}(\tau')  v^*_{k_2}(\tau'') v^*_{k_3}(\tau'')u^{*}_{p_2}(\tau'')u^{*}_{p_3}(\tau'')\notag \\
	&+2\, \text{perms}.
\end{align}
Similarly, Eq.~\eqref {2a2} is simplified as
\begin{align}\label{2a,2s}
	&\left\langle h^{s_1}_{\mathbf k_1}(\tau) h^{s_2}_{\mathbf k_2}(\tau)h^{s_3}_{\mathbf k_3}(\tau) \right\rangle_{2a,2}\notag\\ =&
	-\frac{1}{2}	(2\pi)^3\delta\left(\mathbf k_1+ \mathbf k_2+\mathbf k_3\right) \int^\tau_{\tau_0^*}d\tau' a(\tau')^2  \int^\tau_{\tau_0}d\tau''    a(\tau'')^2 
	\notag \\
	&\times \prod_{A=2}^3\left(\int \frac{d^3p_A}{(2\pi)^{3}}\right) (2\pi)^3\delta\left(\sum_{A=2}^3 \mathbf p_A-\mathbf k_1\right) 
	\notag\\
	&\times   e_{ij}^{s_1*}(\hat k_1) p_{2i}p_{3j}    e_{km}^{s_2*}(\hat k_2) e_{ml}^{s_3*}(\hat k_3) p_{2k}p_{3l}  
 v^*_{k_2}(\tau) v^*_{k_3}(\tau) v_{k_1}(\tau)
 \notag\\
 &\times
   v_{k_2}(\tau')v_{k_3}(\tau') u_{p_2}(\tau') u_{p_3}(\tau')  v^*_{k_1}(\tau'') u^{*}_{p_2}(\tau'')u^{*}_{p_3}(\tau'')\notag\\
	&+2\, \text{perms}.
\end{align}
Note that after exchanging $\tau'$ and $\tau''$, Eq.~\eqref {2a,1s} and Eq.~\eqref {2a,2s} are conjugate to each other. Therefore,
\begin{align}
	\left\langle h^{s_1}_{\mathbf k_1}(\tau) h^{s_2}_{\mathbf k_2}(\tau)h^{s_3}_{\mathbf k_3}(\tau) \right\rangle_{2a}=2{\rm Re}\left\langle h^{s_1}_{\mathbf k_1}(\tau) h^{s_2}_{\mathbf k_2}(\tau)h^{s_3}_{\mathbf k_3}(\tau) \right\rangle_{{2a},1}.
\end{align}
Then, we deal with the polarization tensor and momentum integral parts. Due to the conservation of momentum and without losing generality, we can fix all $k_i$ in the $(x,z)$ plane. Such vector triangles can be constructed as	
\begin{align}\label{kcord}
	\mathbf{k}_1&=k_1(0,0,1),\notag\\ \mathbf{k_2}&=k_2(\sin\theta,0,\cos\theta),\notag\\ \mathbf{k}_3&=k_3(\sin\phi,0,\cos\phi),
\end{align} 
where $\sin\theta=\lambda/2k_1k_2$, $\cos\theta=(k_3^2-k_1^2-k_2^2)/2k_1k_2$, $\sin\phi=-\lambda/2k_1k_3$, $ \cos\phi=(k_2^2-k_3^2-k_1^2)/2k_1k_3 $, with $\lambda=\sqrt{2k_1^2k_2^2+2k_2^2k_3^2+2k_3^2k_1^2-k_1^4-k_2^4-k_3^4}$. Polarization tensors are as follows
\begin{align}
	e_{ij}^{s_1}(\hat k_1) \equiv \frac{1}{2}\begin{pmatrix}
		1 & is_1 & 0 \\
		is_1 & -1 & 0  \\
		0  & 0 & 0
	\end{pmatrix},
\end{align}
\begin{align}
	e_{ij}^{s_2}(\hat k_2) \equiv \frac{1}{2}\begin{pmatrix}
		\cos^2\theta & -is_2\cos\theta & -is_2\sin\theta\cos\theta \\
		is_2\cos\theta & -1 & -is_2\sin\theta  \\
		-is_2\sin\theta \cos\theta &  -is_2\sin\theta & \sin^2\theta
	\end{pmatrix},
\end{align}
\begin{align}
	e_{ij}^{s_3}(\hat k_3) \equiv \frac{1}{2}\begin{pmatrix}
		\cos^2\phi & -is_3\cos\phi & is_3\sin\phi\cos\phi \\
		-is_3\cos\phi & -1 & -is_3\sin\phi  \\
		-is_3\sin\phi \cos\phi &  -is_3\sin\phi & \sin^2\phi
	\end{pmatrix},
\end{align}
  Because of the independence of  the angle in Eq.~\eqref{2a,1s}, we have
  \begin{align}\label{angular}
  	\int \frac{d^3p_2d^3p_3}{(2\pi)^6}(2\pi)^3\delta(\mathbf k_1-\mathbf p_2-\mathbf p_3)\notag\\ =\frac{1}{(2\pi)^2k_1}\int_0^{\infty} d p_2 \int_{|p_2-k_1|}^{p_2+k_1} dp_3 p_2p_3 .
  \end{align}
Note that $\mathbf{p_A}=p_A(\sin\theta_A\cos\phi_A,\sin\theta_A\sin\phi_A,\cos\theta_A)$, combining Eq.~\eqref{2a,1s} and Eqs.~\eqref{kcord}-\eqref{angular}, we obtain
\begin{align}\label{2a,11}
	&\int \frac{d^3p_2d^3p_3}{(2\pi)^6}(2\pi)^3\delta( \mathbf p_2+\mathbf p_3-\mathbf k_1) e_{ij}^{s_1*}(\hat k_1) p_{2i}p_{3j}e_{km}^{s_2*}(\hat k_2)\notag\\&\times e_{ml}^{s_3*}(\hat k_3) (p_{2k}p_{3l}+p_{3k}p_{2l}) f(p_2,p_3)\notag\\
	=&\int_0^{\infty} d p_2 \int_{|p_2-k_1|}^{p_2+k_1} dp_3  {w}^{s_1s_2s_3}_{k_1} f(p_2,p_3).
\end{align}
where
\begin{align}
{w}^{s_1s_2s_3}_{k_1}=	\frac{ A_{k_1k_2k_3}^2k_T^2p_2p_3A_{k_1p_2p_3}^4}{2\pi^2k_1^7k_2^2k_3^2}.
\end{align}
$A_{k_1p_2p_3}$ refers to the area of the triangle enclosed by $k_1$, $p_2$ and $p_3$. 

Then we apply the mode function Eq.~\eqref{evoldel} and Eq.~\eqref{hevoldel} to perform the momentum integration. The calculation is straightforward and the result (for +++ polarization) is
\begin{align}
	& \left\langle h^{+}_{\mathbf k_1}(\tau) h^{+}_{\mathbf k_2}(\tau)h^{+}_{\mathbf k_3}(\tau) \right\rangle_{2a}'
\notag
=
-\frac{1}{2}\left(\frac{H}{M_{\rm p}}\right)^6\Theta_{2-\tilde{k}_1}	\bar{w}^{+++}_{k_1} \notag\\
& \times \;  \frac{1}{k_1^3k_2^3k_3^3 k_*^4} \int^0_{x_0}dx'  \int^0_{x_0}dx''
\mathcal{F}(\tilde{k}_1,x')\mathcal{G}(\tilde{k}_2,\tilde{k}_3,x'') \nonumber \\
  &+\;2\,\text{perms},
\end{align}
which is the same as Eq.~\eqref{h2a}. Note that the definition of ${w}^{s_1s_2s_3}_{k_1}$ is different from $\bar{w}^{s_1s_2s_3}_{k_1}$in Eq.~\eqref{h2a}. We omit the derivation of $\langle \mathcal{O}\rangle_{2b}$ due to the similarity of calculations.

Now we consider the one-loop contribution from the triangle diagram. Taking $\langle \mathcal{O}\rangle_{3a}$ as an example,
\begin{align}
	&\left\langle h^{s_1}_{\mathbf k_1}(\tau) h^{s_2}_{\mathbf k_2}(\tau)h^{s_3}_{\mathbf k_3}(\tau) \right\rangle_{3a}
	\notag \\
	=&
	\frac{1}{4} {\rm Im}\int^\tau_{\tau_0^*}d\tau' a^2(\tau')\int^\tau_{\tau_0}d\tau''   a^2(\tau'') \int^{\tau''}_{\tau_0}d\tau'''a^2(\tau''')\notag\\ &\prod_{A=1}^9\left(\int \frac{d^3p_A}{(2\pi)^{3}}\right)
	\notag 
	(2\pi)^9\delta\left(\sum_{A=1}^3 \mathbf p_A\right)\delta\left(\sum_{A=4}^6 \mathbf p_A\right) 
    \delta\left(\sum_{A=7}^9 \mathbf p_A\right)  
 \notag \\
	&\times\sum_{s,s_4,s_7} e_{ij}^{s}(\hat p_1) p_{2i}p_{3j}    e_{kl}^{s_4}(\hat p_4)  p_{5k}p_{6l}  e_{mn}^{s_7}(\hat p_7) p_{8m}p_{9n} 
	\notag \\
	&	\times  \langle 0| h^{s}_{\mathbf p_1}(\tau')h^{s_1}_{\mathbf k_1}(\tau)h^{s_2}_{{\mathbf k_2}}(\tau)h^{s_3}_{{\mathbf k_3}}(\tau) h^{s_4}_{\mathbf p_4}(\tau'') h^{s_7}_{\mathbf p_7}(\tau''')|0\rangle
	\notag \\
	&\times \langle 0| \delta \chi_{\mathbf p_2}(\tau')  \delta \chi_{\mathbf p_3} (\tau')  \delta \chi_{\mathbf p_5}(\tau'')  \delta \chi_{\mathbf p_6}(\tau'')\delta \chi_{\mathbf p_8}(\tau''')  \delta \chi_{\mathbf p_9}(\tau''')	|0\rangle
	,
\end{align}
Because the scalar field perturbation is amplified only at $k=k_*$, 
\begin{widetext}
\begin{multline}
	\left\langle h^{s_1}_{\mathbf k_1}(\tau) h^{s_2}_{\mathbf k_2}(\tau)h^{s_3}_{\mathbf k_3}(\tau) \right\rangle_{3a} = 
	2 \text{Im}\delta\left(\mathbf k_1+ \mathbf k_2+\mathbf k_3\right)p^3_*\int^\tau_{\tau_0^*}d\tau' a^2(\tau')\int^\tau_{\tau_0}d\tau''   a^2(\tau'') \int^{\tau''}_{\tau_0}d\tau'''a^2(\tau''')\\ \int_0^{\infty} \dd p_3 \, p_3^2 \, \int_{-1}^1 \dd \cos\theta  \, \int_0^{2 \pi} \dd \phi \,
	e_{ij}^{s_1*}(\hat k_1) p_{2i}p_{3j}    e_{kl}^{s_2*}(\hat k_2)p_{2k}p_{2l}   e_{mn}^{s_3*}(\hat k_3) p_{3m}p_{3n}	v^*_{k_1}(\tau) v_{k_2}(\tau) v_{k_3}(\tau)  v_{k_1}(\tau')\\ u^2_{p_*}(\tau')    |u_{p_*}|^2(\tau'') (u^{*}_{p_*})^2(\tau''')\left(v^*_{k_2}(\tau'')v^*_{k_3}(\tau''')+v^*_{k_3}(\tau'')v^*_{k_2}(\tau''')\right)\\\delta\big(p_3-p_*\big)\delta\big(|\vp_3-\vk_1|-p_*\big)\delta\big(|\vp_3+\vk_3|-p_*\big)+2\, \text{perms}.
\end{multline} 
\end{widetext}
we integrate out  momentum and find that the integral has support at the two points
\begin{equation}
	 \vp_3  = k_1 \left( \frac{-k_1^2+k_2^2+k_3^2}{8  \, A_{k_1k_2k_3}} ,\,
	\pm \frac{\sqrt{16 A_{k_1k_2k_3}^2  k_\star^2 - k_1^2 k_2^2 k_3^2}}{4  A_{k_1k_2k_3}  k_1 }	,\, \frac{1}{2}
	\right).
	\label{p1-2}
\end{equation}
\begin{figure}[t!]
		\includegraphics[width=0.4\textwidth]{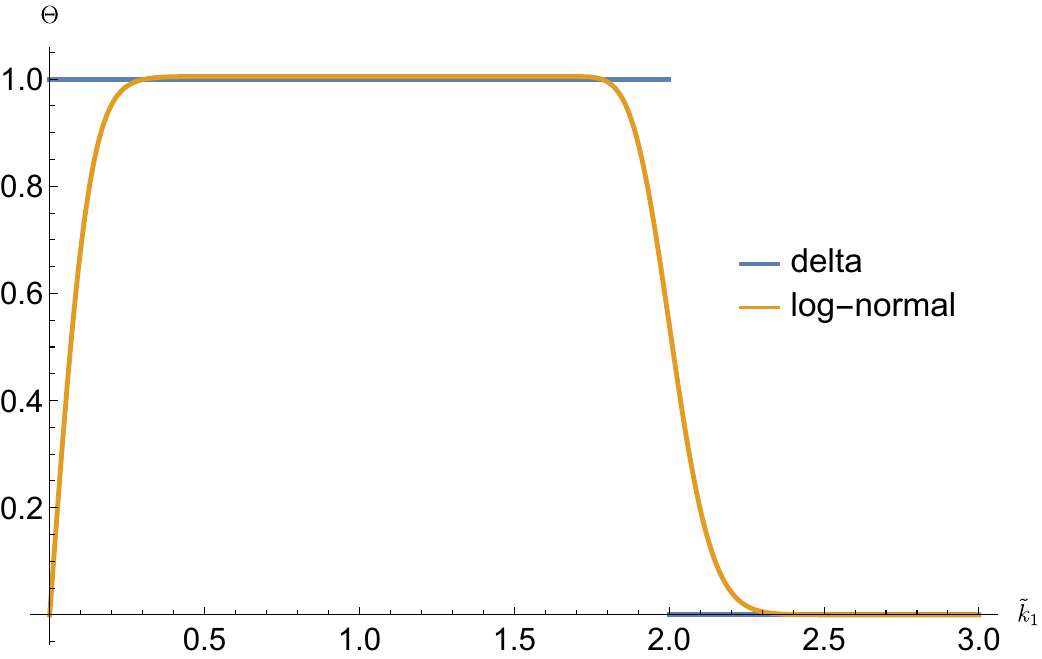}
		\caption{Step function corresponding to delta distribution and log-normal distribution, respectively.}
		\label{theta}
	\end{figure}
The following calculation is similar to the bubble diagram, so we are not prepared to elaborate on details (so are $\left\langle \mathcal{O} \right\rangle_{3b}$). We can see that the above derivation is lengthy and complex. A more elegant and systematic method to compute the correlation functions is the diagrammatic method based on path integral, which provides an equivalent description to canonical in-in formalism \cite{Chen:2017ryl}.

\section{Shape functions under different momentum distributions}\label{B}
 In this appendix, we show the shape functions from the bubble diagram in both delta distribution and log-normal distribution. 

In log-normal case, from Eq.~\eqref{log-normal}, we obtain \cite{Ota:2022xni}
\begin{align}
	f^{\rm LN}(p_2,p_3) =\frac{e^{-\frac{[\ln (p_2/p_*)]^2+[\ln (p_3/p_*)]^2}{2\Delta^2}}}{2\pi \Delta^2} f(p_*,p_*).\label{fln}
\end{align}
we can integrate out one of the momenta, and the remaining integral gives a step function characterized by width $\Delta$
\begin{align}
	\Theta^\Delta_{2-\tilde{k}_1 }\equiv  \frac{e^{ \Delta ^2/2}}{2}\left[
\text{erf}\left(\frac{ \Delta ^2-\ln \left(| 1-\tilde{k}_1| \right)}{\sqrt{2} \Delta }\right)
\right.
\notag \\
\left.-\text{erf}\left(\frac{ \Delta ^2-\ln \left(1+\tilde{k}_1\right)}{\sqrt{2} \Delta }
\right)\right]. 
\end{align}
\begin{figure}[t!]
		\includegraphics[width=0.35\textwidth]{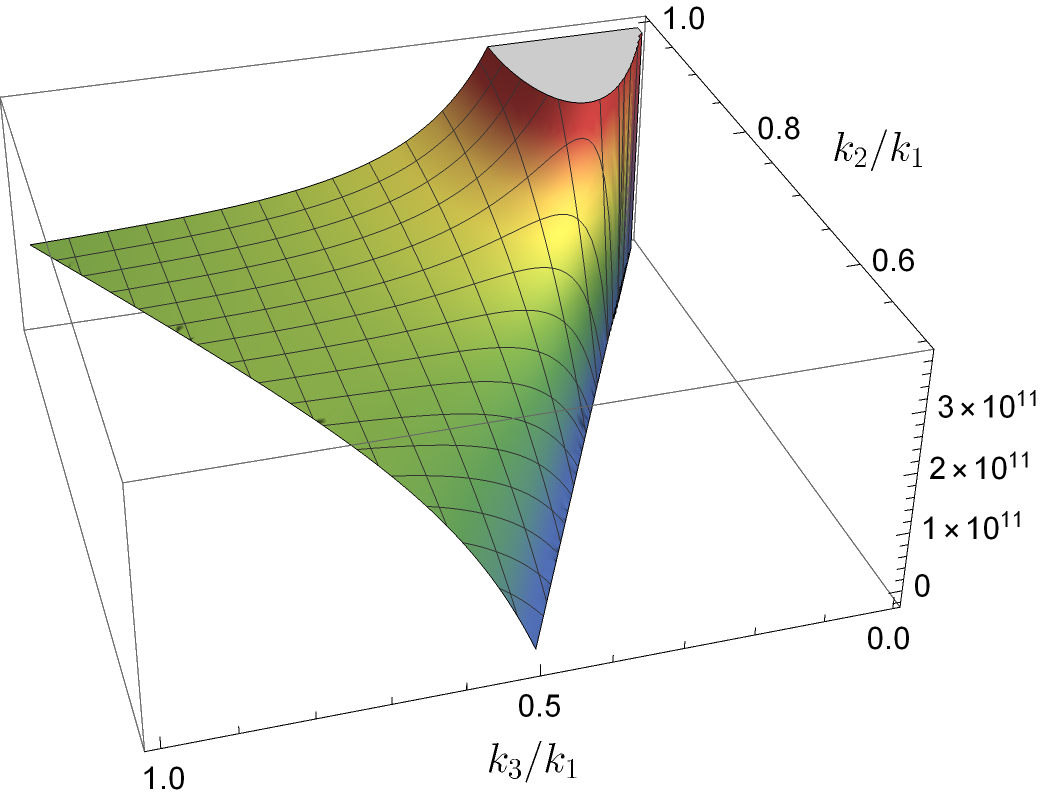}
        \includegraphics[width=0.35\textwidth]{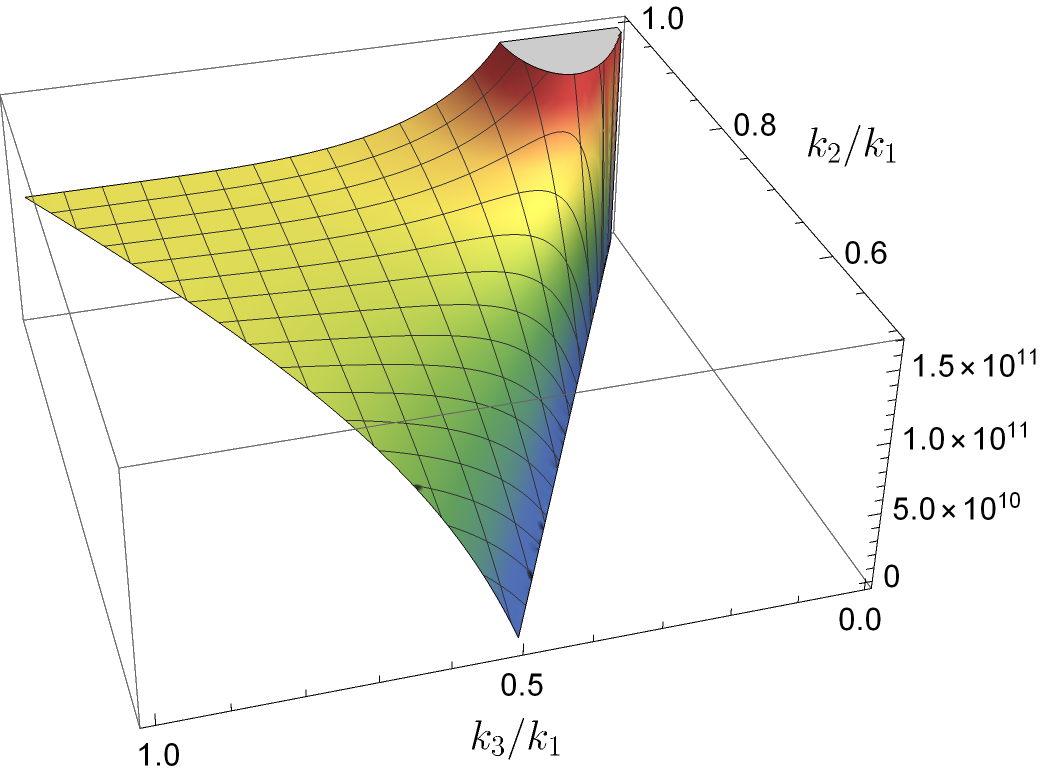}
		\caption{Rescaled shape functions corresponding to different momentum distributions for the $+$+$+$ polarization from the bubble diagram, where we choose $k_1/k_*=0.1$ and $\Delta=0.1$. Top: delta distribution. Down: log-normal distribution.}
		\label{bub}
	\end{figure}
So the main difference between delta and log-normal distribution is $\Theta_{2-\tilde{k}_1 }\to\Theta^\Delta_{2-\tilde{k}_1 }$.
Step functions corresponding to delta distribution and log-normal distribution are illustrated in Fig.  \ref{theta}. We find that $\Theta^\Delta_{2-\tilde{k}_1 }$ deviates significantly  from 1 when k approaches zero. Actually, it behaves as a linear function of $\tilde{k}_1$. The same conclusion applies to the other two polarization configurations. The shape functions under two different momentum distributions are shown in Fig.~\ref{bub}. We can see that the peak value of the shape function in log-normal distribution is smaller than one in the delta case.  Meanwhile, the rescaled shape function corresponding to delta distribution $\propto\tilde{k}_1^{-4}$, thus, the configuration of the tensor bispectrum does not change with the momentum distribution, and the squeezed configuration clearly dominates.

	\bibliographystyle{apsrev4-1}
	\bibliography{bispectrum}
	
\end{document}